\documentclass[journal, letterpaper, twocolumn]{IEEEtran}

\usepackage{graphicx}
\usepackage{color}
\usepackage{epstopdf}
\usepackage{amsmath}
\usepackage{amsbsy}
\usepackage{amssymb}
\usepackage{amsthm}
\usepackage{caption}
\usepackage{subcaption}
\usepackage{epsfig}
\usepackage{psfrag}
\usepackage{booktabs}

\newcounter{tempeqncnt}
\newcommand{\ba}{\boldsymbol{a}}
\newcommand{\bi}{\boldsymbol{i}}
\newcommand{\bo}{\boldsymbol{o}}
\newcommand{\bb}{\boldsymbol{b}}

\newcommand{\bn}{\boldsymbol{n}}

\newcommand{\bx}{\boldsymbol{x}}
\newcommand{\by}{\boldsymbol{y}}
\newcommand{\bzero}{\boldsymbol{0}}

\newcommand{\bE}{\boldsymbol{E}}
\newcommand{\bI}{\mathbf{I}}
\newcommand{\bH}{\boldsymbol{H}}
\newcommand{\bG}{\boldsymbol{G}}
\newcommand{\bS}{\boldsymbol{S}}
\newcommand{\bW}{\boldsymbol{W}}
\newcommand{\bV}{\boldsymbol{V}}

\newcommand{\bU}{\boldsymbol{U}}
\newcommand{\bSigma}{\boldsymbol{\Sigma}}

\newcommand{\complex}[1]{\mathbb{C}^{#1}}
\newcommand{\real}[1]{\mathbb{R}^{#1}}
\newcommand{\calS}{\mathcal{S}}
\newcommand{\calC}{\mathcal{C}}
\newcommand{\calN}{\mathcal{N}}

\newcommand{\E}{\textnormal{\textsf{E}}}
\newcommand{\T}{\textnormal{\textsf{T}}}
\renewcommand{\H}{\textnormal{\textsf{H}}}

\newcommand{\tr}{\mathrm{tr}}
\newcommand{\brc}[1]{\left( #1 \right)}
\newcommand{\sqbrc}[1]{\left[ #1 \right]}
\newcommand{\figbrc}[1]{\left\{ #1 \right\} }

\newcommand{\norm}[1]{\left\|#1\right\|}

\renewcommand{\vec}{\mathrm{vec}}
\renewcommand{\det}{\mathrm{det}}
\newcommand{\maximize}{\mathrm{maximize}}
\newcommand{\st}{\mathrm{subject\;to}}

\newcommand{\wf}{\mathrm{WF}}
\newcommand{\opt}{\mathrm{opt}}

\begin{document}

\title{Deep-Learning Based Linear Precoding for MIMO Channels with Finite-Alphabet Signaling}

\author{Maksym~A.~Girnyk

\thanks{%
M.~A.~Girnyk is with Ericsson Research, Stockholm, Sweden (e-mail: maksym.girnyk@ericsson.com). 

This a perprint of an accepted paper available at\\ \begin{tt}https://doi.org/10.1016/j.phycom.2021.101402\end{tt}}
}

\maketitle

\begin{abstract}
	This paper studies the problem of linear precoding for multiple-input multiple-output (MIMO) communication channels employing finite-alphabet signaling. Existing solutions typically suffer from high computational complexity due to costly computations of the constellation-constrained mutual information. In contrast to existing works, this paper takes a different path of tackling the MIMO precoding problem. Namely, a data-driven approach, based on deep learning, is proposed. In the offline training phase, a deep neural network learns the optimal solution on a set of MIMO channel matrices. This allows the reduction of the computational complexity of the precoder optimization in the online inference phase. Numerical results demonstrate the efficiency of the proposed solution vis-a-vis existing precoding algorithms in terms of significantly reduced complexity and close-to-optimal performance. 
\end{abstract}

\section{Introduction}
\label{sec:intro}

% MIMO in NR
Increased communication data rates are an inherent requirement of the future 5G-connected world. Multiple-input multiple-output (MIMO) technology is of great help for achieving this goal, providing opportunities for antenna arrays to focus the energy in narrow beams, spatially multiplex data streams or robustify the transmission by means of diversity. Multi-antenna deployments and techniques thus constitute an integral part of the 5G NR standard~\cite{asplund2020advanced}.

% Optimum precoding
In the case of correlated antennas, the achievable data rate can be improved by optimizing the precoder matrix. It is known that the maximum achievable rate is given by the Shannon capacity of a MIMO channel which is achieved through the diagonalization of the channel matrix by means of singular value decomposition (SVD) and subsequent \emph{water filling} (WF) over the parallel non-interfering channels~\cite{telatar1999capacity}. This solution is based on the underlying assumption that Gaussian noise-like signals are used for the transmission. The latter is, however, not the case in practice, where signals are instead selected from a finite-alphabet constellation, e.g., phase shift keying (PSK) or quadrature amplitude modulation (QAM). For such signaling schemes, the compact SVD-based WF solution may not perform well~\cite{lozano2006optimum}. 

% Finite alphabets
An optimal precoding strategy for finite-alphabet signaling is formulated in~\cite{lamarca2009linear} and~\cite{xiao2011globally}. The solution is based on an algorithm, optimizing eigenvectors and singular values of the precoding matrix in an alternate manner. The algorithm does converge to the optimal solution, however, it is computationally expensive due to the necessity to empirically evaluate \emph{constellation-constrained} mutual information (MI) and minimum mean square error (MMSE) matrix at each iteration. 

% Survey of existing work
Various algorithms to simplify the optimization of the precoder have been proposed. In~\cite{ketseoglou2014linear}, the approach of per-group precoding is introduced, performing grouping of multiple input streams and receiving branches after the SVD of the channel matrix. In~\cite{yang2019low}, the precoder optimization is converted into a simpler problem of minimization of a negative exponential function. Other works have proposed methods for simplification of the computation of MI. Thus,~\cite{kozachenko1987sample, singh2003nearest}, propose entropy estimation methods based on computation of the nearest-neighbor distance. In~\cite{zhu2003efficient}, a statistical computation approach, based on an approximation of the distribution of the received signal, is developed, showing reduced computational complexity. A constant-gap lower bound to the MI is derived in~\cite{zeng2011low}, enabling faster MI computation. A method for reducing complexity by using a sphere-decoding based approximation of the entropy of a Gaussian mixture is proposed in~\cite{kim2015entropy}. Furthermore, approximations for the computation of MI and MMSE matrix, based on the Gauss-Hermite quadrature, are proposed in~\cite{ketseoglou2016linear}, leading to a more efficient precoder optimization. An approximation based on the Taylor expansion of the log-term under integral and a subsequent least-squares fitting was proposed in~\cite{zhang2017analytical}, simplifying the computation of MI, while keeping decent accuracy. 

% Alternative with ML
Recently, application of deep learning (DL) techniques for various wireless communications problems have gained increasing attention in literature. For instance, a deep-learning (DL) based end-to-end optimization approach using autoencoders is proposed in~\cite{dorner2017deep}, showing competitive performance. In~\cite{wen2018deep}, a DL-based channel state information (CSI) sensing and representation framework is developed improving CSI reconstruction. Meanwhile, in~\cite{lin2019beamforming}, DL-based hybrid beamforming is proposed, demonstrating strong robustness under imperfect CSI. An algorithm based on deep reinforcement learning for beamforming optimization is proposed in~\cite{lee2020deep}, showing nearly-optimal performance. 

\begin{figure*}[!t]
	% ensure that we have normalsize text
	\normalsize
	% Store the current equation number.
	\setcounter{tempeqncnt}{\value{equation}}
	% Set the equation number to one less than the one
	% desired for the first equation here.
	% The value here will have to changed if equations
	% are added or removed prior to the place these
	% equations are referenced in the main text.
	\setcounter{equation}{5}
	\begin{equation}
	\label{eqn:miDiscrete}
	I\brc{\by; \bx | \bH, \bG} = M \log |\calS| - \frac{1}{|\calS|^M} \sum_{m=1}^{|\calS|^M} \E_{\bn} \figbrc{\log \sum_{k=1}^{|\calS|^M} \exp\brc{-\norm{ \bH\bG\brc{\bx_m-\bx_k} + \bn}^2 - \norm{\bn}^2}}
	\end{equation}	
	% Restore the current equation number.
	\setcounter{equation}{\value{tempeqncnt}}
	\vspace*{-0.5cm}
	\hrulefill	
	% The spacer can be tweaked to stop underfull vboxes.
	\vspace*{4pt}
\end{figure*}

% Filling the gap
In this paper, finite-alphabet precoder optimization is combined with DL to obtain a \emph{data-driven} low-complexity solution for the problem of MIMO precoder optimization. Following the supervised learning methodology, it is proposed to split the optimization into two phases: offline training and online inference. In the former, the computationally heavy optimization is done to train an artificial deep neural network (NN) to learn the mapping between the optimal finite-alphabet precoder and the capacity-achieving WF precoder. In latter phase, this mapping is used to directly compute the finite-alphabet precoder for new channel realizations. Provided numerical results illustrate that the proposed DL-based approach has near-optimal performance with low computational complexity.

\section{Problem Formulation}
\label{sec:systemConfig}

A multi-antenna communication system is mathematically described by a MIMO vector channel. Namely, assuming orthogonal frequency-division multiplexing (OFDM) is used, for a given time/frequency resource element the received signal vector $\by \in \complex{N}$ is modeled as
\begin{equation}
\label{eqn:mimoChannel}
	\by = \bH \bG \bx + \bn,
\end{equation}	
where $\bH \in \complex{N\times M}$ is a matrix consisting of the channel coefficients, $\bn \sim \calC\calN \brc{\bzero_N, \bI_N}$ is the noise vector at the receiver, and $\bx \in \complex{M}$ is the transmitted symbol vector, where each entry is picked from a finite-alphabet constellation $\calS$. It is assumed that $\E\figbrc{\bx}=\bzero_M$ and $\E\figbrc{\bx\bx^{\H}}=\bI_M$. Furthermore, $\bG$ denotes the precoder matrix that adjusts the transmission to the given radio environment.

Assume that channel $\bH$ is perfectly known at both the receiver and transmitter. In practice, this is achieved by means of channel reciprocity in time division duplex (TDD) operation. For instance, in 5G NR, uplink sounding reference signals (SRSs) are sent by a mobile device to a base station, so that the latter estimates the downlink channel matrix $\bH$. Having such an estimate, the base station can optimize the precoder $\bG$. This is in contrast to the frequency division duplex (FDD) case, where explicit CSI feedback is unavailable and the precoder matrix is selected from a codebook of possible precoders, indicated by a precoder matrix indicator (PMI) that is fed back from the device.

The precoder optimization problem is formulated as follows
\begin{equation}
	\begin{aligned}
		\underset{\bG}{\maximize} \quad & I\brc{\by; \bx | \bH, \bG}\\
		\st \quad &  \tr\figbrc{\bG^{\H} \bG } \leq M.
	\end{aligned}
\end{equation}
That is, maximization of the achievable data rate, given by the MI $I\brc{\by; \bx | \bH, \bG}$ between the input and output of the MIMO channel, subject to the total power constraint.

The maximum theoretically achievable rate of a MIMO channel is given by the Shannon capacity. That is, under the condition that the transmit signals are drawn from a Gaussian distribution, i.e., $\bx\sim\calC\calN\brc{\bzero_M, \bI_M}$, the MI reads as\footnote{Note that to have the rate in bit/s/Hz the logarithm is taken with base 2.}
\begin{equation}
\label{eqn:miGaussian}
	I\brc{\by; \bx | \bH, \bG} = \log\det \brc{\bI_N + \bH\bG\bG^{\H}\bH^{\H}}.
\end{equation}
The optimal precoder matrix is given by the WF solution~\cite{telatar1999capacity}
\begin{equation}
\label{eqn:svdPrecoder}
	\bG_{\wf} = \bV_{\bH} \bS_{\bG},
\end{equation}	
where $\bV_{\bH}$ is the matrix consisting of the right singular vectors of the channel (obtained from the SVD of the channel matrix: $\bH=\bU_{\bH} \bSigma_{\bH} \bV_{\bH}^{\H}$), and $\bS_{\bG}$ is a diagonal matrix, whose entries are water-filled according to
\begin{equation}
\label{eqn:diagonal}
	\sqbrc{\bS_{\bG}}_{m, m} \!=\! \sqrt{\!\!\brc{\frac{1}{\nu} - \frac{1}{\sqbrc{\bSigma_{\bH}}_{m, m}}\!}^{\!\!+}}\!\!, \forall m \!\in\! \figbrc{1, \ldots, \min\{M, N\}}\!,
\end{equation}	
where $\brc{\cdot}^{+} = \max\figbrc{\cdot, 0}$, and $\nu$ is chosen so that full power is utilized, i.e., $\tr\figbrc{\bG_{\wf}^{\H} \bG_{\wf} } = M$.

Unfortunately, Shannon's capacity~\eqref{eqn:miGaussian} is hardly achieved in practice---except for the very low signal-to-noise ratio (SNR) operation mode---since it is based on the assumption of purely information-theoretical random Gaussian signaling. Practical communication systems are operating with finite-alphabet signal constellations, such as, e.g., BPSK, QPSK and 16-QAM. The above solution is therefore \emph{not optimal} for practical settings, because a truly optimal solution should aim at maximizing an expression of the actual achievable data rate for finite-alphabet signals instead. The latter is given by the constellation-constrained MI between $\by$ and $\bx$, showed in~\eqref{eqn:miDiscrete} on the top of the page, where $\calS$ represents the set of all points of the given signal constellation, and $\E_{\bn}\figbrc{\cdot}$ stands for the expectation w.r.t. the distribution of the noise vector $\bn$.

When using~\eqref{eqn:miDiscrete}, the optimum solution to the optimization problem above is obtained by the iterative algorithm proposed in~\cite{lamarca2009linear} and~\cite{xiao2011globally}. The corresponding solution does not exhibit a closed-form expression; instead, the algorithm is based on alternations between gradient-descent updates of eigenvectors and singular values of the precoder matrix until convergence (see~\cite[Sec. IV-B]{xiao2011globally}). Clearly, the evaluation of the expression for MI~\eqref{eqn:miDiscrete} requires a large number of arithmetic operations for multi-antenna setups with even moderate numbers of antennas. Moreover, in addition to the computation of MI, one has to evaluate its gradient, given by the MMSE matrix~\cite{palomar2005gradient}
\begin{equation}
\setcounter{equation}{7}
\label{eqn:mmseDiscrete}
	\bE \brc{\bH, \bG} = \E\figbrc{\brc{\bx-\hat{\bx}}\brc{\bx-\hat{\bx}}^{\H}},
\end{equation}	
where $\hat{\bx}=\E\figbrc{\bx|\by}$ is the MMSE estimate of $\bx$. Hence, there are further nested computation loops at each iteration of the optimization algorithm, which increases the computation time even further. This all makes the precoding of~\cite{lamarca2009linear} and~\cite{xiao2011globally} infeasible for real-time operation in multi-antenna systems.

\section{Proposed solution}
\label{sec:proposedSolution}

The idea proposed herein is to use supervised DL to train a model (e.g., a deep NN) on a large dataset of channel observations and pre-comupted (offline) optimal precoders. Then, in the online inference phase, the optimized finite-alphabet precoder is obtained for new channel realizations, yet unseen by the transmitter, exploiting the trained model.

\subsection{Training phase}
In the offline training phase, for each MIMO channel matrix from the dataset, an optimal precoder is computed based on a given modulation scheme and using either true expressions of MI and MMSE matrix, or accurate approximations thereof. 

The DL model learns the mapping between the WF precoder and the optimal precoder for the given finite-alphabet constellation, solving the problem of multiple-output regression. Since it is well-known that NNs are universal function approximators~\cite{castro2000neural}, a deep NN serves as a good candidate for such a learning model.

Practically, let $\bG_{\wf}$ be the SVD-based WF precoder and $\bG_{\opt}$ be the truly optimal finite-alphabet precoder for a channel matrix $\bH$. For each $\bH$ from the available training set, a deep NN with $P$ hidden layers is trained to map their vectorized versions converted to a real-valued representation. That is, the input vector for the NN is given by 
\begin{equation}
\label{eqn:nnInput}
	\bi = [\vec(Re\{\bG_{\wf}\})^{\T}, \vec(Im\{\bG_{\wf}\})^{\T}]^{\T} \in \real{2 M^2},
\end{equation}
while the output vector reads as
\begin{equation}
\label{eqn:nnOutput}
	\bo = [\vec(Re\{\bG_{\opt}\})^{\T}, \vec(Im\{\bG_{\opt}\})^{\T}]^{\T}  \in \real{2 M^2}.
\end{equation}
The trained NN then carries all the information needed to obtain the optimal finite-alphabet precoder from the light-weight WF solution, encoded in a set of weight matrices $\bW_p\in \real{L_{p-1}\times L_{p}}$ and bias vectors $\bb_p \in \real{L_p}$, where $p~\in~\figbrc{1,\ldots,P+1}$.

\subsection{Inference phase}
Once the training is done, in the online inference phase, an optimal precoder is obtained by means of a single forward propagation pass. The operation is of very low complexity, consisting of a number of matrix multiplications and additions. That is, for a deep NN with $P$ hidden layers,
\begin{equation}
\ba_{p} = g_p\brc{\bW_p \ba_{p-1} + \bb_p}, \quad \forall p\in\figbrc{1,\ldots,P+1},
\end{equation}
where $g_p(\cdot)$ is the activation function of layer $p$. The input and output are obtained as $\ba_0=\bi$ and $\bo=\ba_{P+1}$, respectively. The finite-alphabet precoder $\bG$ is obtained by reshaping $\bo$ via an inverse mapping to that of~\eqref{eqn:nnOutput}.

\subsection{Complexity Analysis}
\label{sec:complexity}

It can be seen from~\eqref{eqn:miDiscrete} that the computation of MI consists of a matrix multiplication and three nested loops: two for averaging over the symbol vector and one for averaging over the noise. Assuming, for simplicity sake, classical matrix multiplication\footnote{Note that a number of methods for the acceleration of matrix multiplication have been proposed, see, e.g.,~\cite{legall2014powers} for the fastest method to date.} and dropping all constants and non-dominant terms, one can get a reasonable upper bound for the corresponding complexity $\mathcal{O}\brc{T_{\bn}|\calS|^{2M}NM}$, where $T_{\bn}$ is the number of iterations required to average over the noise. 

Since the computation of MMSE matrix~\eqref{eqn:mmseDiscrete} contains similar operations, although performed in a different order, its asymptotic complexity is the same. Therefore, the computational complexity of the entire optimization algorithm is $\mathcal{O}\brc{T_{\bn}T_{BLS}Q_{alg}|\calS|^{2M}NM}$, where $T_{BLS}$ is the maximum number of backtracking line search iterations required to determine the optimum step size at each gradient update~\cite{xiao2011globally}, and $T_{OL}$ is the maximum number of outer-loop iterations to accommodate the algorithm's convergence. 

For comparison sake, let us consider a couple of existing algorithms that perform slightly worse, but have lower computational complexity. The algorithm proposed by Zhu et al.~\cite{zhu2003efficient} avoids one of the loops for averaging over the input signal vector $\bx$. It can be shown that its complexity is upper-bounded as $\mathcal{O}\brc{T_{\bn}T_{BLS}T_{OL}|\calS|^{M}N^2M}$. Meanwhile, the algorithm proposed by Zeng et al.~\cite{zeng2011low} simplifies the MI computation by avoiding the averaging over the noise vector $\bn$. Hence, its complexity is shown to be $\mathcal{O}\brc{T_{BLS}T_{OL}|\calS|^{2M}NM}$. These algorithms are picked from the list of the references mentioned in Sec.~\ref{sec:intro} due to their superior performance in terms of accuracy and computational complexity.

In contrast to all the above algorithms, the proposed DL-based approach \emph{does not} require any outer-loop iterations at all. Instead, it directly provides a nearly-optimum solution through a series of matrix multiplications and additions within a single forward propagation pass. Therefore, it can be shown that its complexity is given by $\mathcal{O}\brc{M^2\brc{L_1+L_{P}}+\sum_{p=1}^{P}L_pL_{p+1}}$, where $P$ is the number of hidden layers in the deep NN, and $L_p$ is the number of neurons in layer $p$. 

To compare the complexities of the above algorithms, fix all the iteration numbers $T_{\bn}$, $T_{BLS}$, $T_{OL}$, as well as the NN configuration (i.e., the number of hidden layers, $P$), and drop the corresponding constants. Assume, furthermore, that the number of neurons in each hidden layer of the NN is proportional to the size of the input and output layers, $L_0=L_{P+1}=2M^2$. The resulting asymptotic complexity scaling in terms of system parameters is presented in Tab.~\ref{tab:complexity}. It can be seen that the proposed approach significantly outperforms all other algorithms in terms of computational complexity\footnote{Note that the DL solution does not exhibit exponential complexity in $M$.}.

\begin{table}
\centering
\caption{Computational complexity of the algorithms.}
\label{tab:complexity}
\begin{tabular}{lc}
	\toprule
	Algorithm & Complexity\\
	\midrule
	True optimum solution~\cite{xiao2011globally} & $\mathcal{O}\brc{|\calS|^{2M}NM}$\\
	Algorithm of Zhu et al.~\cite{zhu2003efficient} & $\mathcal{O}\brc{|\calS|^{M}N^2M}$\\
	Algorithm of Zeng et al.~\cite{zeng2011low} & $\mathcal{O}\brc{|\calS|^{2M}NM}$\\
	Proposed DL-based approach & $\mathcal{O}\brc{M^4}$\\
	\bottomrule
\end{tabular}
\end{table}

\section{Illustration}
\label{sec:illustration}

To illustrate the idea let us train a deep NN based on a dataset consisting of 7000 generated $2\times 2$  MIMO channels with i.i.d. Rayleigh fading, BPSK modulation (i.e., $\bx_m \in \figbrc{\pm 1}^M$) and various SNR values given by $\rho = 1/N\;\E\{\tr\{\bH\bH^{\H}\}\}$. The trained model is then applied to new channel matrices to directly obtain the approximation of the optimal finite-alphabet precoder, avoiding the outer loop.\footnote{The codes for reproducing these results are available at\\ \begin{tt}https://github.com/girnyk/OptimalPrecodingMimo\end{tt}.}

The NN architecture chosen for the purpose of illustration is a fully-connected feed-forward network with two hidden layers. The input and output layers have size $L_0=L_3=2M^2$, their entries being precoders (WF-based and the truly optimal one) vectorized according to~\eqref{eqn:nnInput} and~\eqref{eqn:nnOutput}, respectively. The hidden layers are chosen to have a size twice as large as the size of the input and output vectors, i.e., $L_1=L_2=4M^2$. 

The training is done by means of the stochastic gradient descent with $1000$ epochs, mini-batch of size $10$ and learning rate $0.005$. The training set constitutes $70\%$ of the entire dataset. The cost function is chosen to be the squared norm of the difference between the output and target vectors. To deal with real-valued entries, all the activation functions are chosen to be $g_p\brc{\cdot}=tanh\brc{\cdot},\;\forall p\in\figbrc{1,\ldots,P+1}$. The weights and biases are initialized according to the Xavier rule~\cite{glorot2010understanding}. 

\begin{figure}[t]
	\centering
	\includegraphics[width=8cm, trim=0.5cm 0cm 1cm 0.5cm, clip=true]{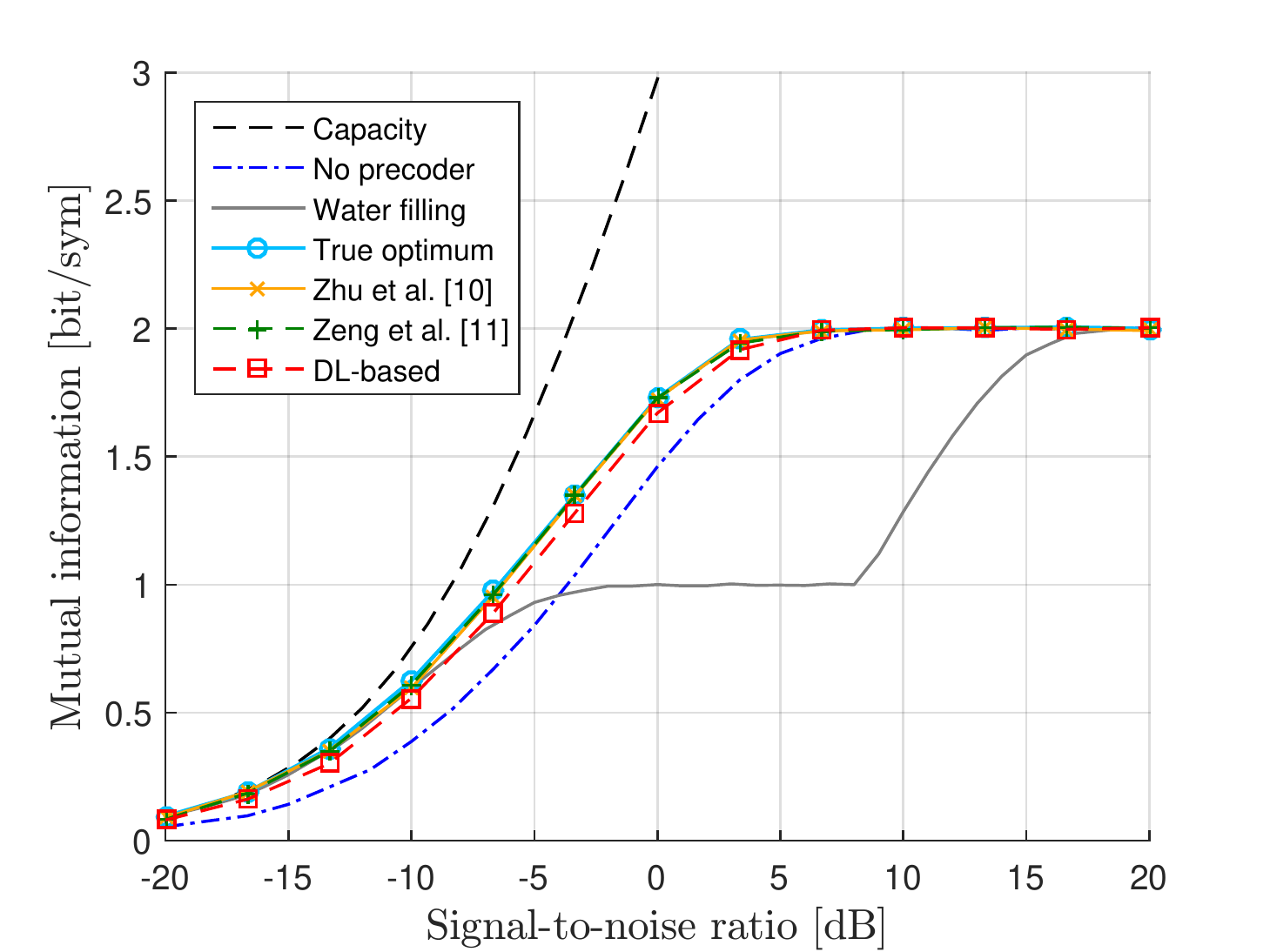}
	\caption{MI between the input and output of $2\times 2$ MIMO channel~\eqref{eqn:channel} with BPSK signaling.}
	\label{fig:mi}
	\vspace{-0.25cm}
\end{figure}

After the training is conducted, for a new channel realization, the WF precoder is computed and vectorized via~\eqref{eqn:nnInput}. It is subsequently inputted to the trained NN. The output of the forward pass through the NN---reshaped inversely to~\eqref{eqn:nnOutput}---provides the precoder matrix to be used as an approximation to the optimal precoder.

Figure~\ref{fig:mi} shows the performance of various solutions in terms of achievable data rate, given by the corresponding MI for the popular channel matrix~\cite{yang2019low, zeng2011low, palomar2005gradient}
\begin{equation}
\label{eqn:channel}
\bH = \sqbrc{\begin{matrix}
	2 & 1\\
	1 & 1
	\end{matrix}},
\end{equation}	
It can be seen that various iterative algorithms, when run for long enough, perform quite close to the optimum. Moreover, the figure illustrates that the NN is able to learn the correspondence between the WF solution and the optimal precoder rather well. It can be observed that the performance of the proposed DL-based solution is quite close-to-optimal, while its execution is very fast (cf. Tab.~\ref{tab:complexity}). Notice also the poor performance of the conventionally ``optimal'' WF solution when it comes to finite-alphabet signaling.

\section{Conclusions}

This paper has presented a novel data-driven approach to the problem of linear precoding for MIMO channels with finite-alphabet inputs. The approach is based on training a deep neural network on a dataset consisting of set of MIMO channel matrices with various signal-to-noise ratios. The features suggested for the training are vectorized precoder matrices based on SVD and water filling, while the suggested labels are vectorized optimal finite-alphabet precoder matrices obtained via~\cite{xiao2011globally}. The learned model provides very fast and reasonably accurate solutions for the precoder optimization problem, while avoiding the time-consuming iterative part inherent to other optimization algorithms. Provided numerical illustration demonstrates the efficiency of the proposed approach on an example of a MIMO channel with Rayleigh fading.

% References
\bibliographystyle{IEEEtran}
\bibliography{references}

\end{document}